\documentclass[twocolumn,pre,aps,10pt,showpacs,superscriptaddress]{revtex4-1}

\usepackage{graphicx}
\usepackage{amssymb,amsfonts,amsmath}
\usepackage{epstopdf}

    \usepackage{multirow}
    
\begin{document}

\title{Competition between monomeric and dimeric crystals in schematic models for globular proteins}

\author{Diana Fusco}
\affiliation{Program in Computational Biology and Bioinformatics, Duke University, Durham, NC 27708}
\affiliation{Department of Chemistry, Duke University, Durham, NC 27708}

\author{Patrick Charbonneau}
\affiliation{Program in Computational Biology and Bioinformatics, Duke University, Durham, NC 27708}
\affiliation{Department of Chemistry, Duke University, Durham, NC 27708}
\affiliation{Department of Physics, Duke University, Durham, NC 27708}

\begin{abstract}
 Advances in experimental techniques and in theoretical models have improved our understanding of protein crystallization. But they have also left open questions regarding the protein phase behavior and self-assembly kinetics, such as why (nearly) identical crystallization conditions sometimes result in the formation of different crystal forms. Here, we develop a patchy particle model with competing sets of patches that provides a microscopic explanation of this phenomenon. We identify different regimes in which one or two crystal forms can coexist with a low-density fluid. Using analytical approximations, we extend our findings to different crystal phases, providing a general framework for treating protein crystallization when multiple crystal forms compete. Our results also suggest different experimental routes for targeting a specific crystal form, and for reducing the dynamical competition between the two forms, thus facilitating protein crystal assembly.
\end{abstract}
 
\maketitle

\section{Introduction}

Crystallizing biomolecules is of central importance for determining their three-dimensional structure through X-ray or neutron diffraction~\cite{PDB,mcpherson:1999}, but remains notoriously difficult to achieve~\cite{khafizov:2014}.
Fortunately, increasing the number of new structures deposited in public databases~\cite{PDB} enriches our understanding of effective crystallization screens and strategies that can be used with the vast majority of biomolecules that still resist assembly~\cite{fusco:2014}. Yet even elementary structural analyses reveal blind spots in our materials comprehension. For instance, among the 80K Protein Data Bank (PDB~\cite{PDB}) deposited structures obtained through X-ray crystallography, 45\% come from monomeric structures, 43\% come from homomers, and 64\% of these homomers result from dimeric assembly. Dimer formation may thus be an important aspect of crystal formation, yet has thus far been mostly neglected~\cite{liu:2012}. %we understand that monomeric and formation may be an important route to protein assembly common assembly units. 
%This observation further raises interesting questions about the possible competition between the two during the crystallization process.

%This observation further raises interesting questions about the possible competition between the two during the crystallization process.

A related feature is that many proteins crystallize in more than one crystal form, with some instances, such as lysozyme, resulting in tens of different unit cells~\cite{Zhang:1995,kohn:2010}. This diversity is partially caused by the variety in crystallization conditions. Different cosolute, pH levels, and salt concentrations can tilt the scale toward different protein-protein interaction mechanisms, leading to the assembly of distinct crystal forms. A very high salt concentration, for instance, strengthens hydrophobic interactions and screens electrostatic ones. 
Yet even under the same solution conditions, and thus presumably similar effective protein-protein interactions, different crystals are sometimes found to assemble~\cite{dixon:1992,faber:1990,mcree:1990,elgersma:1992}. 
This phenomenon has been observed in at least three different experimental contexts: (i) by changing the crystallization temperature~\cite{elgersma:1992}, (ii) by changing the initial protein concentration~\cite{elgersma:1992}, (iii) by letting the crystallization experiment run longer~\cite{dixon:1992,faber:1990,mcree:1990,elgersma:1992}. This last effect is particularly interesting because studies have found that the crystal with the slower growth rate can typically be resolved at a higher resolution; it is less likely to incorporate defects and is thus often preferable, especially for neutron diffraction~\cite{myles:2006}. 

A microscopic understanding of these experimental observations would help target the desirable crystal phase and increase its growth speed and reliability. Studies of small molecules have revealed how atomistic conformational changes result in metastable solid polymorphisms with different nucleation rates~\cite{chung:2009}. In larger and more rigid biomolecules, such as globular proteins, an alternative cause to such phenomenon is the presence of competing crystal contacts. 
In this scenario, one possible mechanism for crystal competition is the formation of protein dimers, which occurs when a given region on the protein surface strongly interacts with the same region on a different chain. Under these circumstances,  a crystal of dimers, which satisfies the specific dimeric interaction, may also compete with a crystal of monomers, which does not carry the dimeric interaction but may be more efficiently packed. Given the reported abundance of dimeric and monomeric crystals, this scenario offers a promising starting point for understanding the role of competition in protein crystal assembly.

From a physical viewpoint, identifying the solution conditions leading to the formation of protein crystals is akin to determining the protein solution phase diagram. In typical experimental setups, a protein is crystallized by super-saturating a low concentration protein solution at constant temperature~\cite{mcpherson:1999}. The crystal that nucleates is thus expected to be the thermodynamically stable form, leaving behind a solution depleted in proteins.
 Early experimental characterization revealed an analogy between the phase behavior of globular proteins and that of colloidal particles with short range interactions~\cite{rosenbaum:1996,Wolde1997}, which both often display a metastable critical point below the crystal solubility line. In these systems, successful crystallization is typically achieved in the region intermediate between the solubility line -above which the solution is stable- and the liquid-liquid critical point -below which the system precipitates into amorphous materials~\cite{charbonneau:2007,Lu2008,Fortini2008}. 
The observation that even spherical and rigid globular proteins are characterized by directional interactions (these interactions are notably responsible for the low packing fraction of protein crystals compared to atomic solids), however, soon led to the replacement of spherical symmetry with \textit{patchiness}~\cite{lomakin:1999,Bianchi2011,fusco:2013a,fusco:2013b,Haxton2012}. In patchy models, a protein is described as a spherical particle decorated with attractive patches that mimic the solution-mediated directional protein-protein interactions driving crystal assembly. Properly parameterized, these models can achieve near-quantitative agreement with experimental phase diagrams of simple proteins~\cite{lomakin:1999,fusco:2013b}. 
%These models have also revealed the important role of interaction anisotropy in the interplay between percolation and crystal assembly~\cite{fusco:2013a}.
 Thus far, however, patchy models have assumed that only patches leading to a single crystal form are present on the protein surface. To take account of competing crystal forms, a dual set of complementary patches that correspond to distinct crystal symmetries has to be included.

%Comparison between experimental data and theoretical models has shown that globular proteins can be described as colloidal particles with short-range directional interactions~\cite{lomakin:1999,Bianchi2011,fusco:2013a,fusco:2013b,Haxton2012}. Under this description, successful crystallization can be achieved between the solubility line, above which the solution is stable, and the liquid-liquid critical point, well below which the system precipitates into amorphous materials~\cite{charbonneau:2007,Lu2008,Fortini2008}. %Despite the success of these models, several protein phase behaviors remain somewhat difficult to rationalize. Some proteins, for instance, such as sickle hemoglobin, exhibit a peculiar inverted solubility, in which saturation increases with temperature rather then decreasing~\cite{ross:1977}, supposedly because of specific protein-solvent interactions~\cite{gunton:2005}. 

In this study, we design a patchy particle model for proteins with competing interactions that can result in both monomeric and dimeric crystal forms. Using numerical simulations, we characterize the model's phase diagram under different interaction parameters
 and test whether it can explain the experimental observation that different crystal forms can assemble depending on (i) crystallization temperature, (ii) initial protein concentration, and (iii) experimental time.  

\section{Methods}

In the following section, we describe the schematic protein model, summarize the details of the simulation techniques, and present analytical approximations that can be used to extend the present analysis to different crystal lattices. 

\subsection{Model}\label{sec:model}

\begin{figure}[tbh]
\begin{center}
\includegraphics[width=0.4\textwidth]{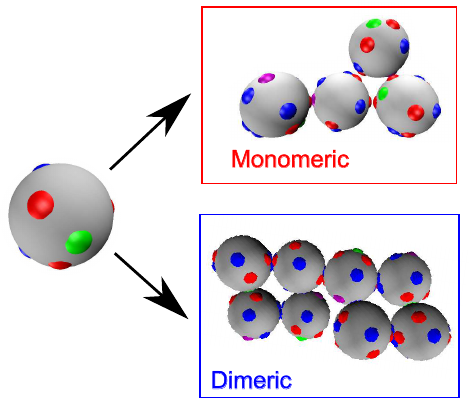}
\caption{Representation of the patchy particle and the two crystals unit cells. The green patch is the dimeric interaction ($D$), the red patches are the crystal contacts of the monomeric unit cell ($m$) and the blue patches are the crystal contacts of the dimeric unit cell ($d$). The purple patch represents patch 2 in Table~\ref{tab:model}, which is a shared crystal contact between the two crystal forms. }\label{fig:model}
\end{center}
\end{figure}

We adopt a schematic patchy particle model for proteins with a distribution of orientational interactions chosen such that the model can form both monomeric and dimeric crystals (Fig.~\ref{fig:model}). 
%These models have been widely used to understand protein assembly~\cite{lomakin:1999, Bianchi2011,fusco:2013a,fusco:2013b,Haxton2012}, and near-quantitative agreement between schematic and the experimental phase diagram has been demonstrated for several globular proteins~\cite{lomakin:1999,fusco:2013b}. 
Each protein is represented by a hard sphere, which models the overall steric repulsion between two proteins, decorated by square-well attractive patches that represent the protein-protein interactions at crystal contact. The solvent contribution is taken to be effective and is thus directly integrated into the patch-patch interaction potential. Because the crystal symmetry P$2_12_12_1$ is the most common for both monomeric and dimeric crystals in the PDB~\cite{PDB}, we assign patch positions according to protein crystals that already have this symmetry. The monomeric crystal patches are chosen to be the same as those used in a previous study of the rubredoxin crystal (PDB: 1BRF)~\cite{fusco:2013b}; the dimeric crystal patches follow the crystal symmetry of the yeast Myo5 SH3 domain (PDB: 1ZUY), whose chain length and crystal density are similar to those of rubredoxin. This choice guarantees that the two crystal forms of the schematic model have comparable number density, although the dimeric crystal happens to be slightly denser. Note that the model does not correspond to a specific protein,  but should be taken as prototype for monomeric--dimeric crystal competition (Fig.~\ref{fig:model}).

%\textbf{Motivate that these interactions neglect charged repultion (crystallization is often at high salt concentration), as well as dispersion attraction -- motivate?.}

Each particles carries a set $\Gamma$ of $n=13$ patches whose pair interactions are depicted in Figure~\ref{fig:model}. Particles 1 and 2, whose centers are a distance $r_{12}$ apart, interact through a pair potential
\begin{equation}
\phi(r_{12},\Omega_1,\Omega_2)=\phi_{\mathrm{HS}}(r_{12})+\sum _{i,j=1}^n\phi_{i,j}(r_{12},\Omega_1,\Omega_2),
\end{equation}
where $\Omega_1$ and $\Omega_2$ are Euler angles describing the orientation of the two particles. The hard-sphere (HS) potential captures the volume exclusion up to a diameter $\sigma$
\begin{equation}
\phi_{\mathrm{HS}}=\left\{
\begin{array}{cc}
\infty&r\leq\sigma\\
0&r>\sigma,
\end{array}\right .
\end{equation}
and patch-patch interactions are the product of a radial and of an angular component
\begin{equation}
 \phi_{i,j}(r_{12},\Omega_1,\Omega_2)=\psi_{i,j}(r_{12})\omega_{i,j}(\Omega_1,\Omega_2).
 \end{equation}
 The radial component depends on the patch type and the inter-particle distance
\begin{equation}
\psi_{i,j}=\left\{
\begin{array}{cc}
-\varepsilon_{i,j}&r\leq\lambda_{i,j}\hbox{ and } p_{i,j}=1\\
0&\mathrm{otherwise},
\end{array}\right .
\end{equation}
where $p_{i,j}$ takes value 1 if patch $i$ interacts with $j$ has reported in Table~\ref{tab:model} and 0 otherwise, whereas $\lambda_{i,j}$ and $\varepsilon_{i,j}$ are the square-well interaction range and strength, respectively. The angular part guarantees that patches only interact when facing each other
\begin{equation}
\omega_{i,j}(\Omega_1,\Omega_2)=\left\{
\begin{array}{cc}
1&\theta_{1,i}\leq\delta_{i} \hbox{ and } \theta_{2,j}\leq\delta_{j}\\
0&\hbox{otherwise}
\end{array}\right .,
\end{equation}
where $\theta_{1,i}$ is the angle between vector $\mathbf{r_{12}}$ and patch $i$ vector on particle 1, $\theta_{2,j}$ is the angle between $\mathbf{r_{21}}=-\mathbf{r_{12}}$ and patch $j$ vector on particle 2, and $\delta_i$ and $\delta_j$ are the semi-angular widths of the patches. The choice of interaction potential implicitly assumes that long-range electrostatic repulsion can be ignored and that no isotropic depletion forces are at play. These assumptions are reasonable for almost 50\% of successful crystallization experiments, in which the relatively high salt concentration screens long-range electrostatics and no depletion agent, such as poly-ethylene glycol, is used~\cite{Charles2006}.  % \textbf{Make a note about the absence/coarse-graining of solvent maybe at the beginning.}

In the following, we simplify the model by assuming that the patch width and interaction range are equal for all patches, using $\delta_i=\mathrm{acos(0.99)}$ and $\lambda_{i,j}=1.2\sigma$ as typical values for protein-protein interactions~\cite{fusco:2013b}. Three types of interaction energies are considered: $\varepsilon_D$ for the patch that holds the dimer together, i.e., the dimeric patch, and $\varepsilon_m$ and $\varepsilon_d$ for the patches corresponding to the crystal contacts of the monomeric and the dimeric crystals, respectively. From the P$2_12_12_1$ lattice geometry, we obtain a lattice energy per particle $e_m=-3\varepsilon_m$ for the monomeric crystal and $e_d=-(\varepsilon_D+5\varepsilon_d)/2$ for the dimeric crystal. Note that although the assumption that all crystal contacts are identical is known not to be generally true, it nonetheless remains qualitatively robust in the limit of small interaction heterogeneity~\cite{fusco:2013a}. Without loss of generality, we adopt reduced units with length being given in units of $\sigma$, and energy and inverse temperature $\beta=1/k_BT$, where $k_B$ is the Boltzmann constant, in units of $\varepsilon_D$. 

\begin{figure}[tbh]
\begin{center}
\includegraphics[width=0.4\textwidth]{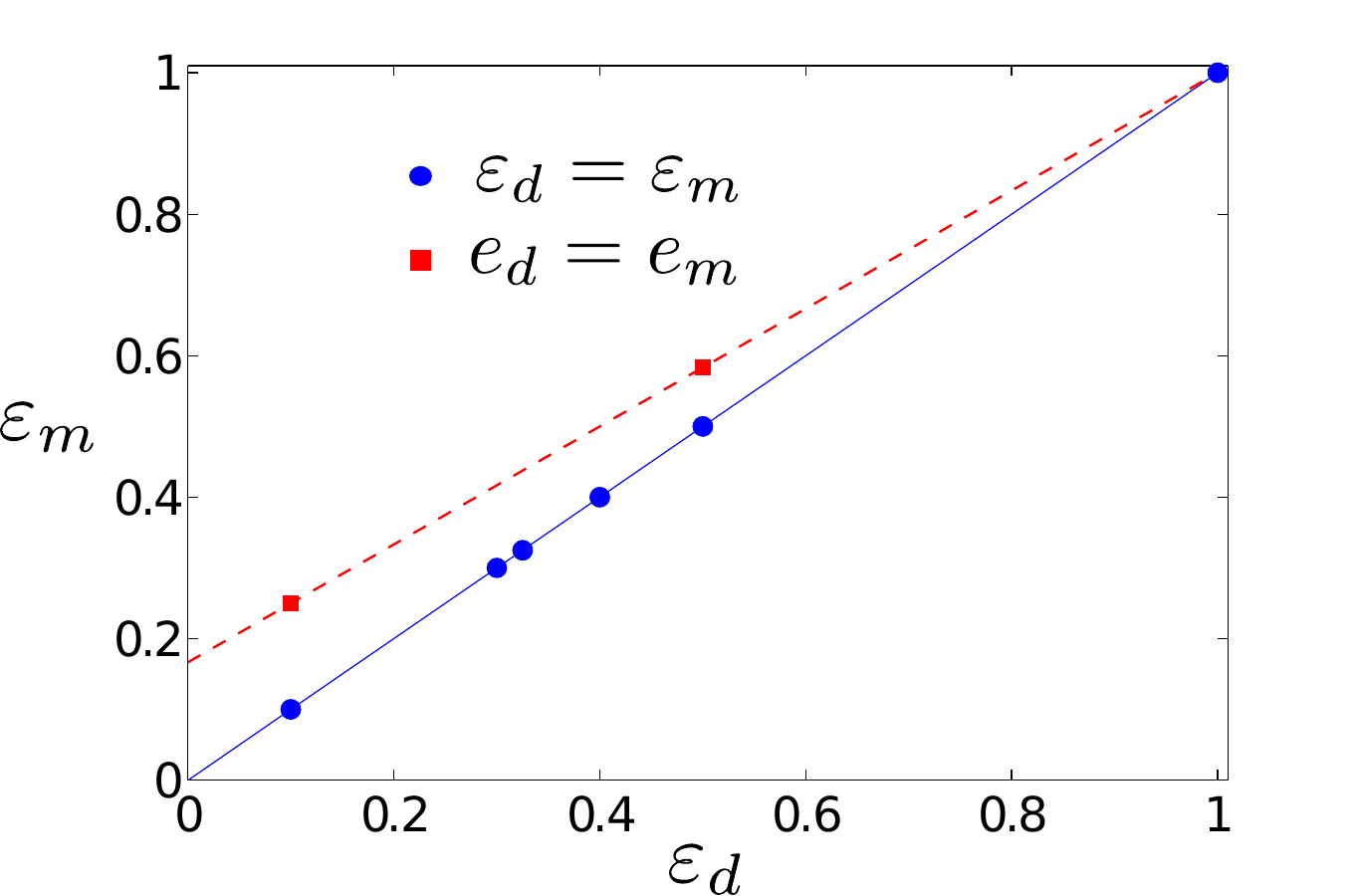}
\caption{Combination of crystal contact strengths explored in this paper. The blue solid line represents the case in which both crystals have identical crystal contact strength, the red dashed line when the energy per particle in the two crystals is the same. The symbols identify the specific parameter sets investigated in simulations.}\label{fig:par_space}
\end{center}
\end{figure}

Our choice of model accounts for systems in which the dimeric patch is stronger than the other, less specific interactions. It explores the regimes in which the dimeric patch strength suffices to control crystal assembly compared to different combinations of $\varepsilon_d$ and $\varepsilon_m$. In particular, we consider the case $\varepsilon_d=\varepsilon_m$, which corresponds to identical crystal contacts for the two crystal forms, and the case $e_d=e_m$ ($\varepsilon_m=\frac{1+5\varepsilon_d}{6}$), which corresponds to an identical lattice energy for the two crystal forms (Fig.~\ref{fig:par_space}). Table~\ref{tab:model} reports the specific positions of the patches on the sphere and their interaction energy pairing.

\begin{table}[tbh]
\begin{center}
\begin{tabular}{ccccc}
\textbf{number}&$\mathbf{\theta}$&$\mathbf{\phi}$&\textbf{interacting pair}&\textbf{type}\\
\hline
1&1.7427&1.2357&1&$D$\\
2&1.0745&5.0987&3/13&$d$/$m$\\
3&1.7424&5.0467&2&$d$\\
4&1.5708&0&5&$d$\\
5&1.5708&3.1415&4&$d$\\
6&0.3032&2.5358&6&$d$\\
7&3.0992&6.2572&8&$d$\\
8&2.7789&4.5241&7&$d$\\
9&1.9858&2.4749&10&$m$\\
10&1.9858&0.6667&9&$m$\\
11&2.7123&3.5542&12&$m$\\
12&0.4293&3.5542&11&$m$\\
13&1.0007&1.1452&2&$m$\\
\end{tabular}
\end{center}
\caption{Patches spherical coordinates (in radiants), interacting patch and types, classified as $D$ for the dimeric patch, $d$ for the dimeric crystal contacts and $m$ for the monomeric crystal contacts. Columns 1 and 4 identify the pairs of interacting patches for which $p_{i,j}=1$}
\label{tab:model}
\end{table}

\subsection{Numerical Simulations}

Phase diagrams for the model family are computed with the help of special-purpose Monte Carlo (MC) simulation methodologies that closely follow the approach used in Ref.~\citenum{fusco:2013a}. The gas-liquid line of the phase diagram is obtained using the Gibbs ensemble method \cite{Gibbs}, and the critical temperature $T_c$ and density $\rho_c$ are extracted using the law of rectilinear diameters~\cite{frenkel:2001}. For these computations, a system of $N=1000$ particles is simulated for an equilibration run of $2\times10^5$ MC cycles that precedes the production run of $3\times10^5$ MC cycles.
Each MC cycle consists of $N$ particle displacements, $N$ particle rotations, $N/10$ particle swaps between the liquid and the gas box and 2 volume moves.

The solubility line is computed by integrating the Clausius-Clapeyron equation starting from a coexistence point determined using free energy calculations and thermodynamic integration \cite{frenkel:2001,Vega2008}. In this case, the crystal free energy is computed by integrating from an Einstein crystal~\cite{frenkel:2001}, whose free energy is evaluated by a saddle point approximation~\cite{fusco:2013a}, and the fluid free energy is integrated from the ideal gas reference state. Free-energy integration over isotherms or isobars identifies the coexistence points between the fluid and the dimeric crystal, the fluid and the monomeric crystal, and the two crystals. Simulations at constant $N$, pressure $P$, and $T$ are run using $N$=500 particles for the fluid and the monomeric crystal, and $N$=512 particles for the dimeric crystal. In this case, each MC cycle corresponds to $N$ particle displacements, $N$ particle rotations, and 2 volume moves that are isotropic for the fluid and anisotropic for the crystal.

To analyze the fluid phase dynamics, we perform MC simulations with $N$=864 particles, at constant $V$ and $T=T_c$ for $3\times10^6$ MC cycles. Previous studies have shown that MC simulations qualitatively capture the Brownian dynamics of patchy particle fluids similar to those used here~\cite{demichele:2006,sanz:2010,fusco:2013a}. During the course of the simulation we track the number and size of monomeric and dimeric crystallites, as well as the number of dimers.  We classify a particle as being part of a monomeric or dimeric crystallite if, respectively, all its monomeric or dimeric crystal contacts are satisfied. Two crystal particles belong to the same monomeric or dimeric crystallite if they are bonded through a monomeric or a dimeric contact, respectively. The number of dimers is straightforwardly defined as the number of particles whose dimeric patch is satisfied divided by 2. The nucleation barriers are calculated using umbrella sampling in constant $NPT$ MC simulations using the crystal cluster size as order parameters, similarly to Ref.~\citenum{fusco:2013a}.

\subsection{Analytical approximations}

We derive analytical approximations for the fluid and crystal free energy using Wertheim perturbation theory and cell theory, respectively,  in order to generalize the simulation results to a broader set of interaction energy values and different crystal lattices. 

\subsubsection{Wertheim perturbation theory for the fluid}

Wetheim's perturbation theory~\cite{wertheim:1984a,wertheim:1984b} approximates the free energy of a fluid of patchy particles as the HS free energy~\cite{HS} and a bond free energy correction 
\begin{equation}
a_f=a_{\mathrm{HS}}+a_{\mathrm{bond}},
\end{equation}
where
\begin{equation}
\beta a_{\mathrm{bond}}=\sum_{i=1}^n\left(\ln X_i-\frac{X_i}{2}+\frac{1}{2}\right).
\end{equation}
Here $X_i$ is the probability that the particle is not bonded at patch $i$. The chemical potential is then given by
\begin{equation}
\beta\mu_f=\beta a_f+\frac{\beta P}{\rho}=\beta a_{\mathrm{HS}}+\beta a_{\mathrm{bond}}+\frac{\beta P_{\mathrm{HS}}}{\rho}+\frac{\beta P_{\mathrm{bond}}}{\rho},
\end{equation}
where $\rho=N/V$ is the number density and the pressure contribution to bonding is 
\begin{equation}
\beta P_{\mathrm{bond}}=\rho^2\sum_{i=1}^n\left(\frac{\partial X_i}{\partial\rho}\right)\left(\frac{1}{X_i}-\frac{1}{2}\right).
\end{equation}

The value of $X_i$ can be determined by solving the equation of mass-action
\begin{equation}
X_i=\frac{1}{1+\sum_{j=1}^n\rho X_j\Delta_{i,j}}.
\end{equation}
Following the notation of Ref.~\citenum{jackson:1988},  we define $\Delta_{i,j}$ as
\begin{equation}
\Delta_{i,j}=\int g_r(12) f_{i,j}(12)d(12),
\end{equation}
where $d(12)$ denotes an integral over all orientations and separations of two particles, $g_r$ is the radial distribution function of HS fluids and $f_{i,j}=\mathrm{exp}[-\beta\phi(12)]-1$ is the Mayer function. In a short-range Kern-Frenkel-like model~\cite{kern:2003}, such as the one adopted here, $\Delta_{i,j}$ can be approximated by using the contact value of the radial distribution function $g_{1^+}$, such that 
\begin{equation}
\Delta_{i,j}=\pi[1-\mathrm{cos}(\delta)]^2(\lambda-1)g_{1^+}.
\end{equation}

As showed in Ref.~\citenum{jackson:1988}, for patches that interact with a single partner, it follows that
\begin{equation}
X_i=\frac{2}{{1+\sqrt{1+4\rho\Delta_{i,j}}}},
\end{equation}
where $j$ denotes the partner site. In our model, all interactions but three fall into this category. For the other three (2, 3 and 13 in Table~\ref{tab:model}), because patch 2 interacts with both 3 and 13 with different energies, the solution comes from solving 
\begin{eqnarray}
X_2+\rho X_2[\Delta_{2,3}X_3+\Delta_{2,13}X_{13}]&=&1\nonumber\\
X_3+\rho \Delta_{2,3} X_2 X_3&=&1\nonumber\\
X_{13}+\rho\Delta_{2,13} X_3 X_{13}&=&1.\label{eq:wertheim}
\end{eqnarray}
In the special case $\varepsilon_m=\varepsilon_d$, we also have that $\Delta_{2,3}=\Delta_{2,13}$, hence $X_3=X_{13}=\frac{X_2+1}{2}$, which offers an analytical solution
\begin{equation}
X_2=\frac{-(1+\rho\Delta_{2,3})+\sqrt{1+6\rho\Delta_{2,3}+\rho^2\Delta_{2,3}^2}}{2\rho\Delta_{2,3}}.
\end{equation}

\subsubsection{Cell model for crystals}

Previous studies~\cite{Sear1999,fusco:2013a} of the fluid-crystal coexistence of patchy particles approximated the chemical potential of the crystal as
\begin{equation}
\beta\mu_c=\beta e_c-s_c+\beta P/\rho_c\sim\beta e_c-s_{\mathrm{SC}},
\end{equation}
where $e_c$ and $s_c$, indicate the lattice energy per particle and the entropy per particle, respectively. This approximation assumes that $\beta P/\rho$ is small, which is realistic when studying liquid-crystal coexistence at low pressure, and that the entropy of the crystal, whatever its symmetry, can be roughly approximated by that of a simple cubic lattice, $s_{\mathrm{SC}}$. This last assumption, however, breaks down when analyzing coexistence between two crystals, or otherwise the only remaining contribution is the difference in lattice energy. 

\begin{figure}[tbh]
\begin{center}
\includegraphics[width=0.4\textwidth]{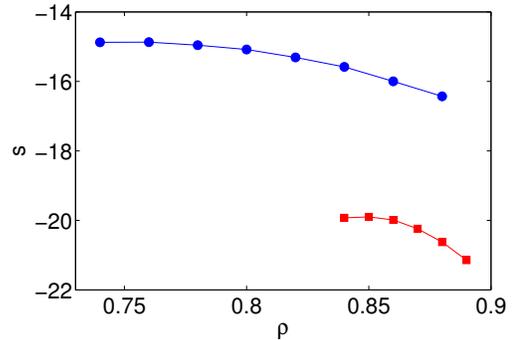}
\caption{Cell-model crystal entropy of the monomeric (blue circles) and dimeric (red squares) lattices as a function of crystal density determined using Monte Carlo integration. The lines are provided as a guide for the eye.}\label{fig:entropy}
\end{center}
\end{figure}

A cell model for a crystal approximates the partition function by the free volume of each particle~\cite{gompper:2007}. Similarly to the HS case, we use a Voronoi tessellation to  divide the crystal in cells $V_i$, each containing a single particle $i$, by construction. The partition function is then obtained by assuming perfect decorrelation of the cells
\begin{eqnarray}
Z&\approx&\left[\int_{V_{i}} d\mathbf{r}_i d\Omega_i \exp(-\beta\sum_{i<j}\phi_{i,j})\right]^N\nonumber\\
&=&\left[\exp(-\beta e)\int_{\hat{V}_i} d\mathbf{r}_i d\Omega_i +\right.\nonumber\\
&&\left.+\int_{V_{i}-\hat{V}_i} d\mathbf{r}_i d\Omega_i \exp(-\beta\sum_{i<j}\phi_{i,j})\right]^N,\nonumber
\end{eqnarray}
where $\hat{V}$ represents the fraction of cell volume in which all square-well interactions between particle $i$ and the surrounding particles are satisfied, and thus depends on the crystal density. The rest of the volume $V_i-\hat{V}_i$ has a much smaller Boltzmann weight and, at low $T$, can be safely ignored. The crystal Helmotz free energy is then
\begin{equation}
a_c=-\frac{1}{\beta}\mathrm{log}Z=e-\frac{1}{\beta}\mathrm{log}\hat{V},\\
\end{equation}
and the crystal entropy $s_c=-\mathrm{log}\hat{V}$~\cite{Sear1999,vega:1998}. Because of the unusual geometry of the crystals considered here and the orientationally-dependent pair-wise potential, we determine $\hat{V}$ by performing a Monte Carlo integration over the unit cell and the angular space as in Ref.~\citenum{vega:1998} (Fig.~\ref{fig:entropy}). 

%\subsection{Analytical approximations}

%In the interesting case of $\varepsilon_m=\varepsilon_d$, it is useful to identify the interaction strength range that corresponds to the different regimes observed in the simulations. To do so, we use analytical approximations to estimate the pressure and the chemical potential of the fluid and the crystals, which can then be used to identity the coexistence lines between two phases.

\section{Results and discussion}

%are analyzed and discussed below. 

In this section, we discuss the analytical and simulation results for two sets of parameters: (i) equal lattice energy ($e_d=e_m$), and (ii) equal crystal contact energy ($\varepsilon_d=\varepsilon_m$). The relatively good agreement between simulation and theory allows us to extend the analytical treatment and to draw more general insights into the role of competition in protein crystal assembly.

%The analysis reveals that in the first case the monomeric crystal is always the only crystal in which the fluid would crystallize. Its slightly lower density makes it more accessible from the fluid phase. In the second case, however, three distinct regimes are identified depending on the crystal contact energy. Above an upper threshold of $\varepsilon_d=\varepsilon_m$, the fluid only crystallizes in the monomeric crystal, below another lower threshold  of $\varepsilon_d=\varepsilon_m$, only the dimeric crystal is stable, and both crystals are stable and depending on the temperature the fluid crystallizes into one or the other.

%\textbf{Write a note to the effect that a protein crystallization experiment mostly increases the density of proteins at fixed $T$.}

%\subsection{Simulation results}

%In this section we report the phase diagrams for different models obtained using Monte Carlo simulations.

\subsection{Case 1: Equal lattice energy}

\begin{figure*}[tbh]
\begin{center}
\includegraphics[width=\textwidth]{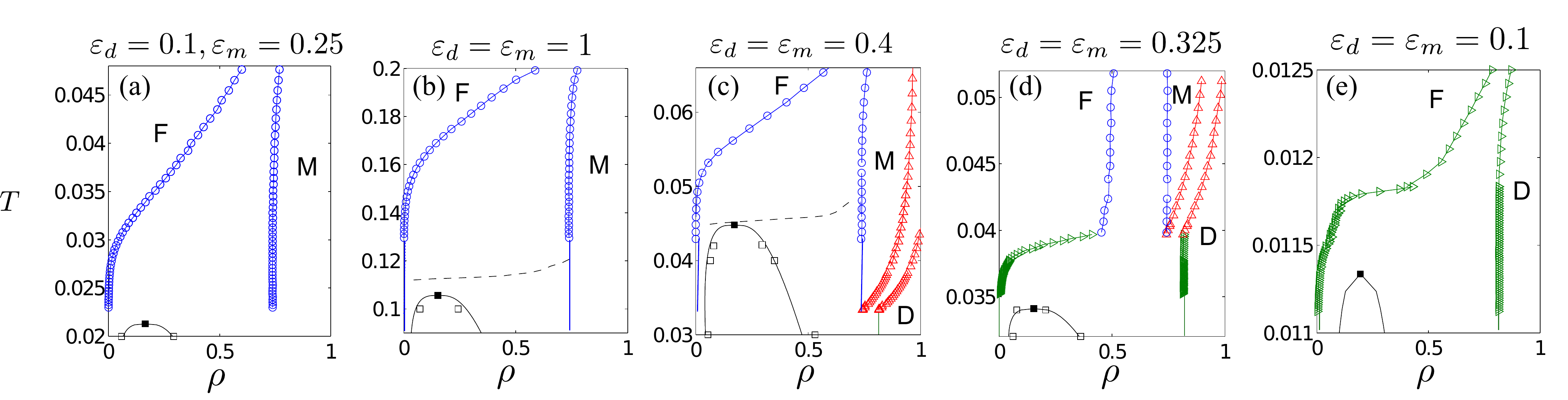}
\caption{Temperature-density phase diagrams for a set of models. (a) $e_d=e_m=0.75$, (b) $\varepsilon_d=\varepsilon_m=1$, (c) $\varepsilon_d=\varepsilon_m=0.4$, (d) $\varepsilon_d=\varepsilon_m=0.325$, and (e) $\varepsilon_d=\varepsilon_m=0.1$. Black squares indicate the liquid-liquid coexistence line and the filled squares the critical point obtained through fitting (black line). Blue circles indicate the fluid--monomeric crystal coexistence, red triangles monomeric-dimeric crystal coexistence, and green right-pointed triangles fluid--dimeric crystal coexistence (lines are guides for the eye). The dashed line represents the metastable fluid-dimeric crystal coexistence line. The letters indicate the stability regions for the fluid (F), the monomeric crystal (M) and the dimeric crystal (D). }\label{fig:PD}
\end{center}
\end{figure*}

In the case where both crystal forms have the same lattice energy, we find that the dimeric crystal phase has no thermodynamic advantage over the monomeric crystal at low-to-medium protein fluid concentrations. Independently of the crystal contact energy, the fluid thermodynamically coexists with the monomeric crystal (Fig.~\ref{fig:PD}). Although stable, the dimeric crystal is only found at very high pressures and $\rho>1$.
%In general, at low temperatures, the crystal with \textit{lower} density will always be closer to the fluid in terms of density and, therefore, the fluid would crystallize into such crystal. More care should be paid, however, if the crystal with higher density has also higher entropy. In such case, the lower density crystal would be stable only up to a maximum temperature (triple point), beyond which the higher density crystal would take over. For sufficiently large entropy differences, richer phase diagrams, similar to those analyzed below, would then emerge.

\begin{figure*}[tbh]
\begin{center}
\includegraphics[width=\textwidth]{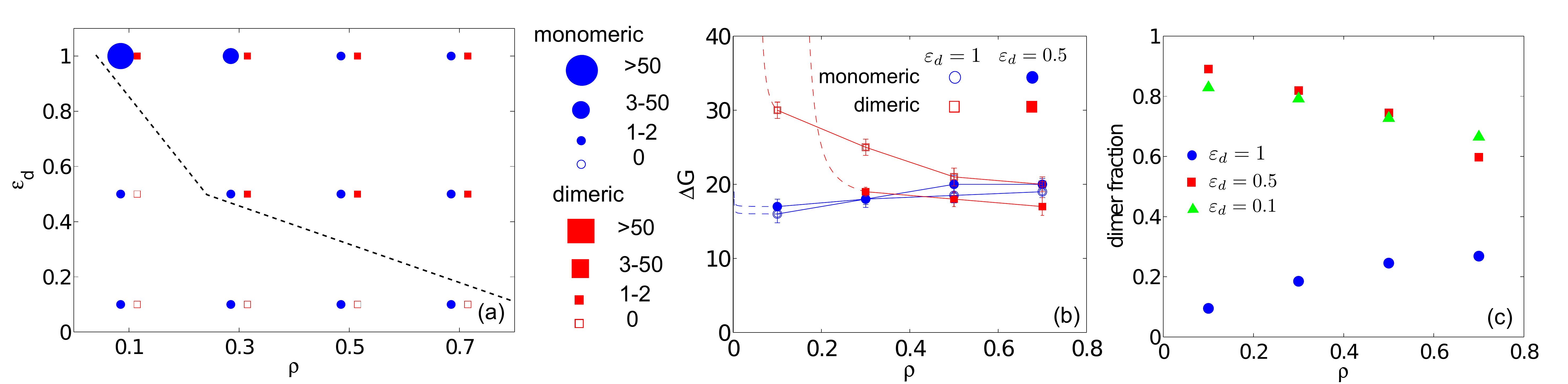}
\caption{Nucleation behavior of the two crystals. (a) Size of monomeric and dimeric crystals as a function of the fluid density and crystal energy $e_d=e_m$ after $3\times10^6$ MC steps at $T_c$. The dashed line is the limit of solubility of the metastable dimeric crystal. (b) Nucleation barriers as a function of fluid density for monomeric (blue circles) and dimeric (red squares) crystals at $\varepsilon_d=1$ (empty symbols) and $\varepsilon_d=0.5$ (filled symbols). Note that barriers for the metastable dimeric phase can only be obtained for densities above the metastable fluid-dimeric crystal line identified in Fig.~\ref{fig:PD} (dashed lines). (c) Steady-state fraction of dimers in the metastable supersaturated fluid.}\label{fig:nucleation}
\end{center}
\end{figure*}

It is important to note, however, that when $\varepsilon_m\approx\varepsilon_d\ll 1$ the fluid is mostly made of dimers. Dimers therefore have to be broken for the monomeric crystal to assemble. Hence, kinetically-controlled assembly could result in the crystal phase that is not the thermodynamically stable one. In order to test if that is the case, we track the type of crystallites in the non-equilibrium fluid for different densities and crystal contact energies. For $\varepsilon_m=\varepsilon_d=1$ at $\rho<0.4$, monomeric crystallites grow into large clusters, while the dimeric crystals remain small (Fig.~\ref{fig:nucleation}). At $\rho>0.4$, however, only a few small monomeric crystallites are observed, while the number of dimeric crystallites increases. Remarkably, although the fluid supersaturation increases with density, crystallization of the thermodynamically preferred crystal becomes increasingly dynamically suppressed, even though the system is still far from the glassy regime. In light of this observation, it is reasonable to hypothesize that as the fluid density increases, crowding enhances the occurrence of interactions that are incompatible with the monomeric crystal, i.e., dimeric crystal contacts and patches. In other words, the fluid-monomeric-crystal interfacial free energy likely increases, which increases the barrier to crystal nucleation. %Monomeric and dimeric interactions compete with each other and no crystal spontaneously grows over the course of the simulation, although both of them are thermodynamically more stable than the fluid phase. 

The dimeric patch inhibits monomeric crystal assembly even more so when $\varepsilon_d<1$, as the energetics of dimer formation grows increasingly favored (Fig.~\ref{fig:nucleation}). As $\varepsilon_d$ is lowered, the number of dimers indeed steadily increases and no spontaneous monomeric crystal growth is observed at any density, over the course of the simulation. 
At $\varepsilon_d=0.5$ and $\rho\gtrsim 0.3$, the dimeric crystal is more stable than the fluid, as confirmed by the growth (melt) of a dimeric crystal seed above (below) this density, and the near constant satisfaction of the dimeric patch suggests that the metastable dimeric crystal may be kinetically accessible. Although the nucleation of the metastable phase is clearly observed in Case 2 (see below), specialized rare-event sampling methodologies are needed to conclusively settle this issue. Calculations of the nucleation barriers for the two crystals at different densities supports the kinetic control hypothesis (Fig.~\ref{fig:nucleation}). The height of the barrier increases with density for the monomeric crystal and decreases for the dimeric crystals, in spite of the fact that the drive to crystallize $\Delta\mu$ steadily increases. The interfacial free energy between the fluid and the monomeric crystal thus rapidly increases with density. Above $\rho=0.5$, the free energy barrier to forming a dimeric crystal is even lower than the monomeric one, suggesting that the former likely precipitates earlier. Yet, at much longer times, the dimeric crystal should eventually transform into the thermodynamically-preferred monomeric crystal through a (much slower) crystal reorganization.
The formation of dimeric crystals under kinetic control can therefore explain two key experimental observations: different initial protein concentrations leading to different crystals, depending on whether the concentration is above or below the metastable dimeric crystal solubility line; and different experimental waiting time leading to different crystal structures.

It is also interesting to note that for $\varepsilon_d<1$, the fraction of dimers in the fluid decreases with increasing fluid density. At low density, clusters of size larger than two are rare due to their entropic cost, and therefore the strongest (dimeric) patch dominates clustering; by contrast, at high densities, multiple particles come in contact and the weaker patches more easily participate in the formation of aggregates, without necessarily satisfying all dimeric interactions. Increasing temperature, however, leaves only dimers in the metastable fluid at all densities, because the effective stickiness of weaker patches lowers and the dimeric patch becomes the dominating interaction type.
Note that a similar phenomenon has been reported for patchy particles with a single, narrow patch~\cite{munao:2013}. In these models, simple oligomers are observed at low densities, but larger clusters form in high-density fluid. In our case, the dimeric patch is too narrow to allow for multiple contacts, but the other (weaker) patches play that role in its stead.

%\textbf{I did not think before how interesting that could be. How difficult would it be to do a cheap kinetic study? Just by increasing the density of the fluid and seeing what spontaneously forms? IF POSSIBLE, PLEASE START IT NOW. Start at low $\rho$ and $T$ just above $T_c$, and see if there is a crossover in $\rho$ and/or $\varepsilon$.} Working on it.

%\textbf{Write something to the effect that switching the maximal density of the crystals would give the opposite result.}

\subsection{Case 2: Equal crystal contact energy}

The topology of the phase diagram dramatically changes upon tuning $\varepsilon_d=\varepsilon_m\leq1$ (Fig.~\ref{fig:PD}).
At $\varepsilon_d=\varepsilon_m=1$ the only stable crystal is the monomeric crystal and the phase diagram is qualitatively similar to the one discussed in the previous section. Under thermodynamic control, a low-density protein solution would necessarily crystallize in the monomeric form.
Decreasing the crystal contact strength allows for the dimeric crystal form to appear at $\rho<1$, but up to $\varepsilon_d=\varepsilon_m\approx0.4$ the monomeric crystal remains the thermodynamically stable phase. Yet, kinetically, no monomeric crystallite spontaneously grows from the fluid. Instead, for $\rho\geq0.1$, dimeric crystallites quickly grow over the course of the simulation. The lattice that first assembles is thus clearly kinetically-controlled. Similarly to what is observed in Case 1, however, crystal growth is slower at higher density, because crowding enables pair interactions that interfere with crystal assembly.

Because the lattice energy of the dimeric crystal is lower than that of the monomeric form, we expect the region of stability of the latter to disappear at very low $T$, where the entropic contribution to the free energy is small. The accessibility of this phase depends on the relative position of the fluid--dimeric crystal--monomeric crystal triple point with respect to the gas-liquid critical point. The phase diagram at $\varepsilon_d=\varepsilon_m=0.325$ illustrates this situation. In this case, the triple point temperature $T_t$ is higher than $T_c$, and therefore both crystal forms are thermodynamically accessible from the low density fluid: the monomeric crystal at higher temperatures and the dimeric crystal at lower temperatures. Interestingly, a comparable effect has been observed in lysozyme. Under identical solution conmpositions, but different temperatures (5$^{\circ}$C vs. 35$^{\circ}$C), lysozyme assembles in different crystal forms~\cite{elgersma:1992}. The absence of deposited crystal structures, however, prevents us from determining if a lysozyme dimer is involved in the assembly.
%Experiments have shown that changing the degree of supersaturation, i.e. the ratio between the starting protein concentration and the crystal solubility, by either changing the initial protein concentration or by changing the temperature of crystallization, can affect the resulting crystal form~\cite{elgersma:1992}. The observed temperature effect can be easily explained in the context of a phase diagram such as the one at $\varepsilon_d=\varepsilon_m=0.325$.
%Because different temperatures correspond to different levels of supersaturation with respect to the solubility curve \textbf{try to make this point a bit clearer: both $\rho$ and $T$ correspond to a supersaturation wrt the solubility line}, this result agrees with the experimental observation that super-saturation controls the type of crystal, when multiple crystal forms are reported \textbf{Really? $T$ like supersaturation? If yes, please add references}.

Note that when $\varepsilon_d=\varepsilon_m$ becomes sufficiently low, the triple point temperature increases so much that only very high density fluids could coexist with the monomeric crystal. In actual protein crystallization experiments, reaching such a high protein concentration, however, typically results in denaturation~\cite{vekilov:2002}, hence the patchy particle description breaks down. In this case, the monomeric crystal is thus assumed to be unreachable and the protein can only form a dimeric crystal.

\begin{figure}[tbh]
\begin{center}
\includegraphics[width=0.4\textwidth]{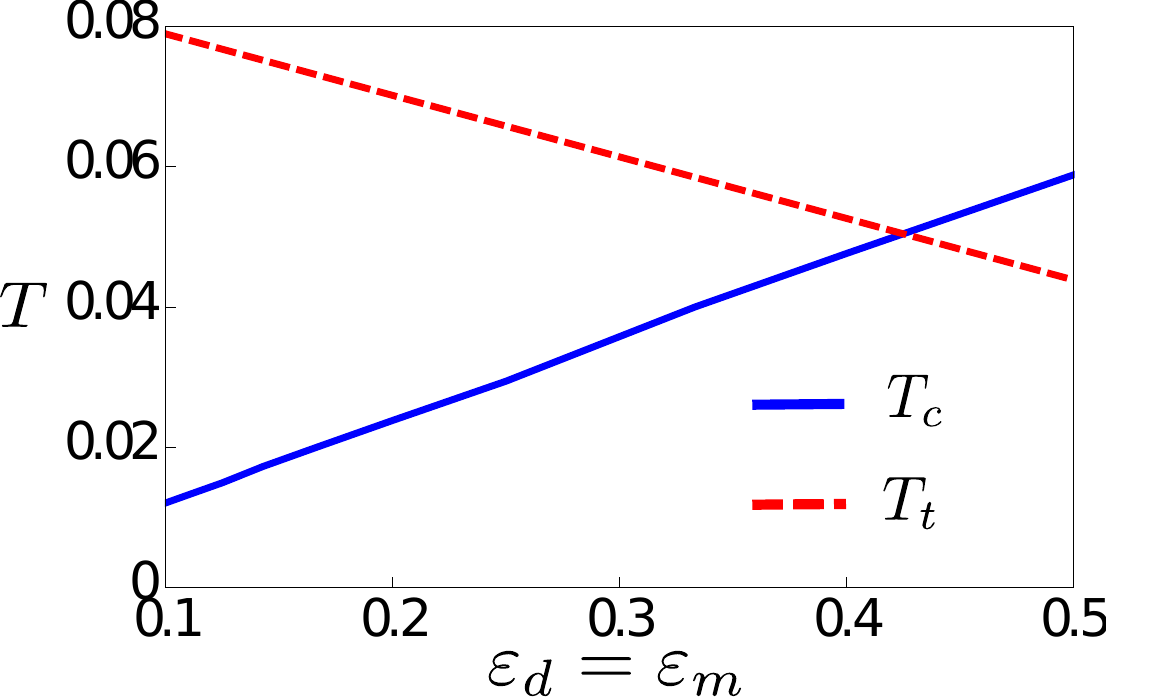}
\caption{Critical temperature $T_c$ (blue solid line) and triple temperature $T_t$ (red dashed line) as a function of $\varepsilon_d=\varepsilon_m$. The crossing point identifies the crossover $\varepsilon^{+}$ above which only the monomeric crystal can be obtained, and below which both monomeric and dimeric crystals are accessible.}\label{fig:tc_tt}
\end{center}
\end{figure}

\begin{figure}[tbh]
\begin{center}
\includegraphics[width=0.4\textwidth]{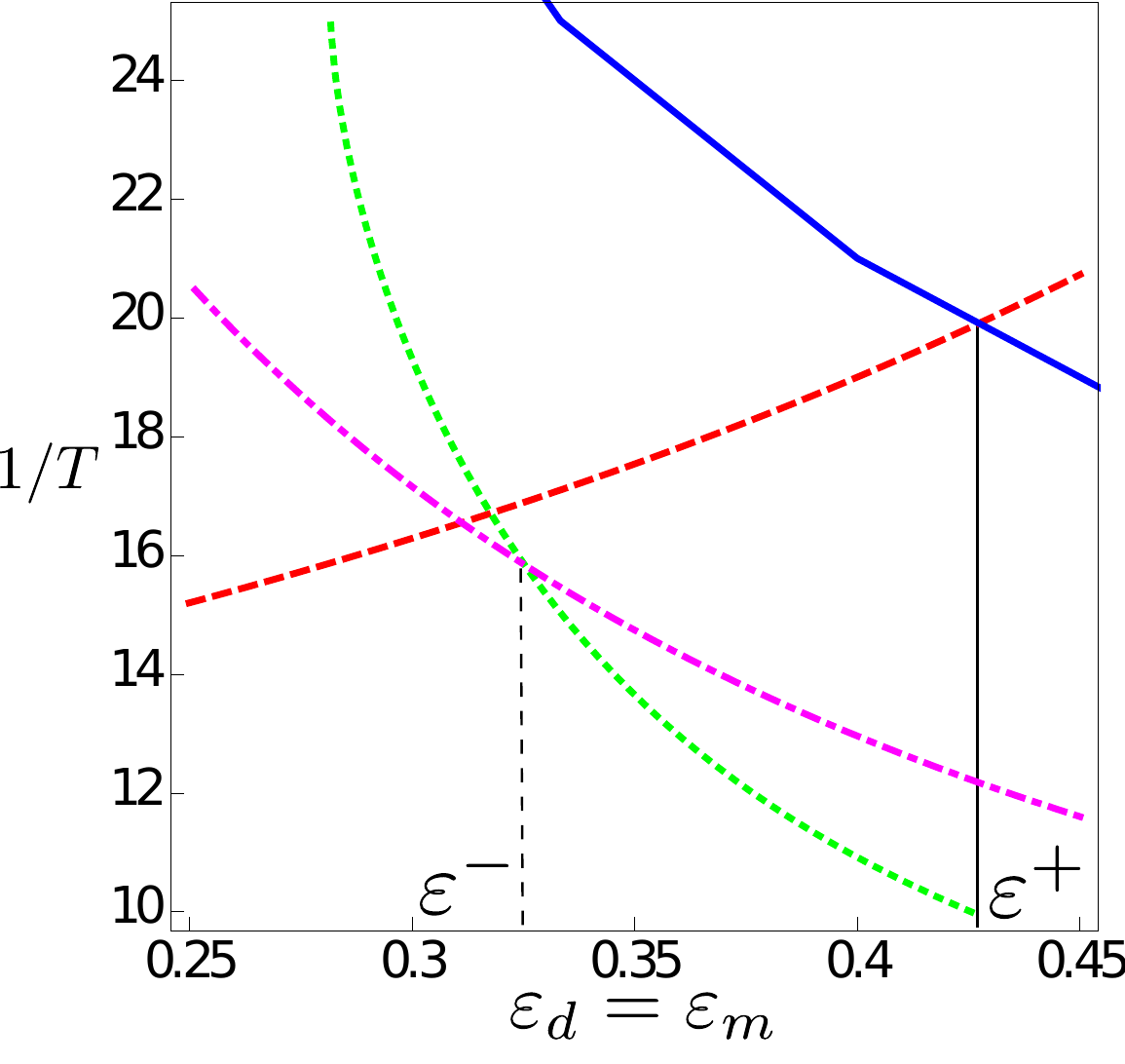}
\caption{Crystal stability regimes. The blue line is the critical temperature, the red dashed line is the coexistence temperature between dimeric and monomeric crystals at the low pressure limit. The green dotted line and the magenta dot-dashed line are the coexistence temperatures between the fluid at $\rho=0.8$ and the monomeric and dimeric crystals, respectively. The vertical solid and dashed black lines identify the upper $\varepsilon^+$ and lower $\varepsilon^-$ thresholds.} \label{fig:threshold}
\end{center}
\end{figure}

Figure~\ref{fig:PD} indicates that in general weakening crystal contacts lowers the critical point and raises the triple point. 
The analytical Wertheim-cell model allows us to extend these simulation results to any $\varepsilon_d=\varepsilon_m$, so as to better capture the thermodynamics of crystal assembly.~\footnote[4]{In agreement with other reports~\cite{vega:1998}, 10\% discrepancy between cell theory and simulations is observed, which explains why at $\varepsilon_d=\varepsilon_m=0.4<\varepsilon^+$ the theory predicts a triple point above the critical point in contrast to the simulation results~(Fig.~\ref{fig:threshold} vs. Fig.~\ref{fig:PD}).}
The case $\varepsilon_d\neq\varepsilon_m$ could also be examined by numerically solving Eq.~(\ref{eq:wertheim}), but is not considered here. %\textbf{Say that the case with different $\varepsilon$ could be solved numerically but is not conidered here} 
Equating the chemical potential and pressure of low- and high-density fluids provides an estimate for the critical temperature~(blue line in Fig.~\ref{fig:tc_tt})~\cite{fusco:2013a}.
Because at low pressure the monomeric--dimeric crystal coexistence $TP$ line is almost vertical, we can safely assume that the triple point between the fluid and the two crystals, i.e., the minimum stability temperature of the monomeric crystal, is found in the low-pressure regime. In this case, equating the free energies of the two crystals provides the triple point temperature $T_t$ (red line in Fig.~\ref{fig:tc_tt})
\begin{equation}
%\beta \mu_d\sim\beta e_d-s_d(\rho_d)&=&\beta e_m-s_m(\rho_m)\sim\beta \mu_m\nonumber\\
\frac{1}{T_t}\sim\frac{s_d(\rho_d)-s_m(\rho_m)}{e_d-e_m}.\label{eq:tt}
\end{equation}
The critical and triple temperatures crossover at a specific value $\varepsilon_d=\varepsilon_m=\varepsilon^{+}$. For $\varepsilon_d=\varepsilon_m>\varepsilon^{+}$, $T_c>T_t$, and thus, in order to crystallize in the dimeric form, densifying a protein fluid would have to cross the metastable gas-liquid coexistence regime. Gelation should then typically render this crystal form inaccessible~\cite{charbonneau:2007,Lu2008,Fortini2008}. For  $\varepsilon_d=\varepsilon_m<\varepsilon^{+}$, the triple point is higher than the critical point. A crystallization experiment at a temperature $T_c<T<T_t$ would result in dimeric crystal formation.
 Comparing the position of the critical and the triple temperature thus identifies an upper threshold $\varepsilon^{+}$ above which the fluid can only crystallize in the monomeric crystal, and below which both crystal forms are attainable by tuning the system temperature.

At low $\varepsilon_d=\varepsilon_m$, the monomeric crystal is only reachable from a very high-density fluid (Fig.~\ref{fig:PD}), which, as argued above, is experimentally inaccessible. A second threshold $\varepsilon^-$ for $\varepsilon_d=\varepsilon_m$, below which only the dimeric crystal coexists with the low density fluid can thus be defined. The value of $\varepsilon^-$ is estimated as the point, where the triple point involves a high-density fluid (arbitrarily set to $\rho=0.8$), which is obtained by equating the pressure and chemical potential of the three phases,
\begin{eqnarray}
e_m-T s_m(\rho_m)+P_f/\rho_m&=& \mu_f\nonumber\\
e_d-T s_d(\rho_d)+P_f/\rho_d&=&\mu_f.\label{eq:tt2}
\end{eqnarray}
Here, $P_f$ and $\mu_f$ are the values for a high-density fluid, and $\rho_m$ ($\rho_d$) are the monomeric (dimeric) crystal coexisting densities. For $\varepsilon_d=\varepsilon_m$, independently of the actual value of the energy, $\rho_d\sim0.89$ and $\rho_m\sim0.82$, which specifies the crystal entropies in Eq.~(\ref{eq:tt2}) (Fig.~\ref{fig:entropy}). 

Figure~\ref{fig:threshold} summarizes the interplay between the critical point, the triple point at low pressure, and the triple point coexisting with a high density fluid. When $\varepsilon_d=\varepsilon_m<\varepsilon^-$, the monomeric crystal is only stable when it coexists with a fluid with $\rho>0.8$, while the dimeric crystal coexists with a low-to-medium density fluid. In the range $\varepsilon^-<\varepsilon_d= \varepsilon_m<\varepsilon^+$, the fluid coexists with both the dimeric (at lower temperatures and densities) and the monomeric (at higher temperatures and densities) crystals. For $\varepsilon_d=\varepsilon_m>\varepsilon^+$, the critical temperature rises above the triple point and the dimeric crystal becomes inaccessible. From this analysis we get that the thermodynamic competition between the two crystal forms occurs over a relatively small range of parameter space. Hence, generally, a single symmetry is thermodynamically stable, despite the presence of several competing patches leading to different crystal forms.  Kinetic control is thus likely to be a more prevalent mechanism for driving crystal contention than outright thermodynamic competition.

\section{Conclusions}

In this work, we have used a simple patchy particle model to study competing crystal forms in the context of protein crystallization, and extended our findings by using analytical approximations from Wertheim and cell theory.
% In particular, we focus on the case in which a protein spontaneously dimerizes because of the presence of a stronger, more specific interaction, which gives rise to a competition between a crystal of dimers, which satisfy the specific interaction but is entropically less favorable, and a crystal of monomers.
Our analysis reveals that tuning the relative strength of crystal-contact and dimeric interactions results in qualitatively different equilibrium phase diagrams, and suggests that certain protein crystal assemblies may be under kinetic control. 
These results provide a microscopic explanation for some of the experimental observations gathered over the years. The existence of a fluid-crystal triple point positioned above the critical point identifies a possible scenario that explains why different crystallization temperatures may result in distinct crystal forms. If this solution behavior is detected experimentally, our analysis suggests that tuning the sample temperature should be an effective strategy to obtain the desired crystal and control its quality. A more invasive alternative would be to mutate amino acids involved in the dimeric patch: strengthening the dimer would widen the stability range of the dimeric crystal, and weakening it would favor the monomeric crystal. Crosslinking the proteins along the dimeric interaction, for instance, would fully suppress the monomeric crystal form. The dimeric crystal would then become the only stable crystal form and larger crystallites should be able to grow. Similar strategies have already been experimentally used to crystallize recalcitrant proteins by encouraging them to dimerize~\cite{Banatao:2006}. A notable example is racemic protein crystallization, where the introduction of a protein mirror image favors pseudo-dimer assembly~\cite{yeates:2012}. Beyond the general structure of the phase diagram, we also show that more subtle experimental observations can be understood by analyzing the dynamics of protein assembly. Changing the crystal form by tuning the initial protein concentration or the experimental time are characteristic of kinetically-controlled assembly. In this context, we find that \emph{decreasing the protein concentration}, and hence the fluid supersaturation, may counter-intuitively \emph{accelerate} crystal assembly. 

% To increase its growth rate at higher concentrations, the dimeric patch, or, in general, the strong interaction that prevents its assembly, has to be weakened, either by mutating it, or by changing the solution conditions. 

%If both crystals have identical energy, the lower density crystal is the one the fluid would crystallize in. Reciprocally, if the crystal contacts of the two crystals are the same, which gives the dimer crystal an energetic advantage, the stable crystal depends on the crystal contact energy and the entropy difference between the two crystals. We identify a lower and an upper threshold of crystal contact energies whose values depend on the crystal entropy. For energies above the upper threshold, only the monomer crystal coexists with the fluid, below the lower threshold only the dimer crystal does, while at intermediate values both crystals are accessible. These results are in agreement with experiments in lysozyme crystallization, which identified different crystal forms depending on the crystallization temperature.

The schematic model introduced here provides a general framework for understanding how crystal lattices with different energies, entropies, and densities assemble.
For example, if the higher-density crystal had a higher entropy, the position of the two crystals in the $T$--$\rho$ phase diagram would be flipped, but the fluid phase would mostly remain unaffected. A different assembly kinetic would then be expected. %The triple point between the crystals and the fluid would then identify the maximum temperature of stability  of the lower entropy crystal rather than the minimum temperature for the higher entropy crystal. 
%Kinetic studies of similar models would also enrich the information obtained from the phase diagram and identify important dynamical features. 
We thus anticipate that future analyses on these richer models for protein assembly will provide better guidance for macromolecular crystallization.

\section{Acknowledgement}
We thank J. Skinner for his gracious support and encouragements over the years. We acknowledge support from National Science Foundation Grant No. NSF DMR-1055586.


\bibliography{dimer}
\bibliographystyle{prsty} %the RSC's .bst file

\end{document}